\documentclass[aps,apl,preprint,showpacs]{revtex4}
\usepackage{graphicx}
\begin{document}

\title{Tailoring exchange bias in half-metallic La$_{2/3}$Sr$_{1/3}$MnO$_{3}$ thin films for spin-valve applications.}

\author{P. K. Muduli, R. C. Budhani}
\affiliation{$^1$Condensed Matter - Low Dimensional Systems
Laboratory, Department of Physics, Indian Institute of Technology,
Kanpur - 208016, India}

\email{rcb@iitk.ac.in}
\date{\today}
\begin{abstract}
\baselineskip=1.5cm

We have utilized the antiferromagnetic nature and
structural/chemical compatibility of
La$_{0.45}$Sr$_{0.55}$MnO$_{3}$ with highly spin polarized
La$_{0.67}$Sr$_{0.33}$MnO$_{3}$ to prepare epitaxial exchange bias
couples. A robust exchange bias (EB) shift of magnetization
hysteresis with associated interfacial exchange energy J $\approx$
0.13 erg/cm$^2$ at 10 K  along with enhanced coercivity are
reported. The EB effect was engineered to bring coercivity
contrast between
 La$_{0.67}$Sr$_{0.33}$MnO$_{3}$ and cobalt films in
La$_{0.45}$Sr$_{0.55}$MnO$_{3}$/La$_{0.67}$Sr$_{0.33}$MnO$_{3}$/SrTiO$_{3}$/Co
magnetic tunnel junctions.
\end{abstract}
\pacs{XX-XX, XX-XX-XX}
%\keywords{}
\maketitle
% start main body here................
%\clearpage \baselineskip=1.5cm
\baselineskip=1.5cm
 Spin-polarized transport in oxide based
half-metals is a topic of intensive research these days because of
its potential for application in spintronics devices. Some of the
most investigated half-metallic oxides include CrO$_2$,
Fe$_3$O$_4$, double perovskites like Sr$_2$FeMoO$_6$ and
manganites La$_{1-x}$Sr$_x$MnO$_3$ (LSMO) for certain doping. In
particular, the manganite with stoichiometry
La$_{0.67}$Sr$_{0.33}$MnO$_{3}$ has been investigated widely for
magnetic tunnel junctions (MTJs)\cite{fert}, for which it is
necessary to have a significant difference in the coercive fields
of the two ferromagnetic electrode in order to have a step-like
switching of magnetization and tunneling conductance. Typically,
exchange bias (EB) phenomenon between a ferromagnet
(FM)/antiferromagnet (AF) bilayer is used to engineer the
coercivity in MTJs. The EB manifests itself as a shift of the
magnetization loop along the field axis by a field H$_{ex}$ and
the coercive field H$_C$ is enhanced. While mechanism of exchange
bias since its discovery 40 years ago by Meiklejohn and
Bean\cite{meikejohn} is still not fully understood, it is widely
believed to arise from the exchange coupling between spins at the
interface between FM and AF layers.

It has been a long standing problem to find a suitable exchange
bias antiferromagnet for La$_{0.67}$Sr$_{0.33}$MnO$_{3}$
(LSMO(FM)), while a positive exchange bias is seen in
La$_{0.67}$Sr$_{0.33}$MnO$_{3}$/SrRuO$_{3}$ bilayer where
SrRuO$_{3}$ is a ferromagnet with T$_{C}$ $\approx$ 150
K\cite{ke}. Interestingly, La$_{1-x}$Sr$_x$MnO$_3$ for x = 0.55 is
an A-type antiferromagnet with pseudo-two-dimensional structure
consisting of double-exchange controlled metallic FM planes
aligned antiferromagnetically along the c-axis. Clearly,
La$_{0.45}$Sr$_{0.55}$MnO$_{3}$ (LSMO(AF))is a potential candidate
for exchange biasing La$_{0.67}$Sr$_{0.33}$MnO$_{3}$  due to its
good lattice and resistivity matching with the latter. In our
previous studies of La$_{0.45}$Sr$_{0.55}$MnO$_{3}$ thin films we
have shown that it can be grown on SrTiO$_3$ (STO) substrate
epitaxially. Such films undergo a N$\acute{e}$el transition at
T$_N$ $\approx$ 220 K\cite{muduli,izumi}.

In this letter we show clear signatures of EB in LSMO(AF)/LSMO(FM)
bilayers. The exchange bias effect has been used to control
coercivity mismatch between LSMO(FM) and cobalt films and in
LSMO(AF)/LSMO(FM)/STO/Co MTJ structures deposited on (100) STO
substrates.

LSMO(AF)(80nm)/LSMO(FM)(50 nm)/STO(3nm) structure was prepared by
pulsed laser deposition (PLD) method on (001) STO. All the layers
were grown at 700$^{0}$C in oxygen pressure of 0.4 mbar. The
details of thin film preparation are reported
elsewhere\cite{muduli,senapati}. Later on a 50 nm thick Co thin
film was deposited at 150 $^{0}$C on some of the
LSMO(AF)/LSMO(FM)/STO structure by e-beam evaporation. The
abbreviation Sample-B and Sample-T has been used in the manuscript
to represent heterostructure
La$_{0.45}$Sr$_{0.55}$MnO$_{3}$/La$_{0.67}$Sr$_{0.33}$MnO$_{3}$/SrTiO$_{3}$
and
La$_{0.45}$Sr$_{0.55}$MnO$_{3}$/La$_{0.67}$Sr$_{0.33}$MnO$_{3}$/SrTiO$_{3}$/Co
respectively. The exchange bias was established through
magnetization measurements carried out in a superconducting
quantum interference device (SQUID) based magnetometer (Quantum
Design MPMS-XL5).

Figure 1(a, b) shows the temperature dependence of field-cooled
(FC) and zero-field-cooled (ZFC) magnetization M(T) measured in a
in-plane field of 1500 Oe for Sample-B (Fig. 1(a)) and Sample-T
(Fig. 1(b)). The magnetic moment of the antiferromagnet LSMO(AF)
(M(5 K) $\approx$ 32 emu/cc)
 is usually an order of magnitude smaller than the moment of
LSMO(FM) (M(5 K) $\approx$ 462 emu/cc)\cite{muduli,senapati}.
Therefore, the signature of AF transition at T$_N$ $\approx$ 220 K
of the 45/55 composition is buried in the strong M(T) response of
the ferromagnetic LSMO(FM) layer. The magnetic moment for the
Sample-T is double compared to the moment of the cobalt-free
sample as seen in Fig. 1(b). This difference can be attributed to
the higher magnetic moment of the Co, which is extracted by
subtracting the moments of Sample-T and Sample-B. The result of
this substraction is shown in the inset of Fig. 1(b). The moment
of the Co film is relatively temperature independent in the range
of 5 to 350 K as seen in the inset. A splitting of the FC and ZFC
branches of M(T) curves can also be seen below 50 K in Fig. 1(a,b)
as indicated by arrows. Such splitting appears to be due to the
onset of exchange interaction between uncompensated spins of
LSMO(FM) and LSMO(AF) at the interface. This can be viewed as a
blocking temperature T$_B$ ($\approx$ 50 K) of the bilayer system.

Magnetization loops at 10 K for Sample-B and Sample-T measured
after cooling in zero field are shown in Fig. 2(a) and 2(b)
respectively. The measurement was done after saturating the sample
at 2000 Oe in-plane field. A shift in the hysteresis loop by
$\approx$ 83 Oe towards the positive field direction is clearly
seen in Fig. 2(a). The coercive (H$_{C}$) and exchange (H$_{ex}$)
fields determined from switching fields H$_{C}^{+}$ and
H$_{C}^{-}$ according to H$_{C}$=(H$_{C}^{+}$-H$_{C}^{-}$)/2 and
H$_{ex}$=(H$_{C}^{+}$+H$_{C}^{-}$)/2 are 60 and 83 Oe
respectively. The shift of the loop seen even on zero-field
cooling (ZFC) indicates an antiferromagnetic coupling between
uncompensated spins at the LSMO(AF)/LSMO(FM)
interface\cite{ke,nogues}.

In order to establish exchange bias effect, Sample B and T are
cooled from 350 to 10 K with an in-plane field of 2000 Oe followed
by measurement of hysteresis loops by sweeping the field over the
range $\pm$ 2000 Oe. A shift in the hysteresis loop for the
bilayer was observed depending on the direction of reference field
used for cooling as shown in Fig. 3 (Panels a $\&$ b). For $\pm$
2000 Oe an exchange bias shift H$_{ex}$$\approx$ $\pm$ 68 Oe is
calculated from the M-H loops. Also the coercive field H$_{C}$ is
enhanced by $\approx$ 26 Oe as compared to the H$_{C}$ of the ZFC
case. The exchange field H$_{ex}$ was also found to decrease
rapidly to zero for temperature above $\approx$ 50 K. The EB
effect seen here is strikingly different from what is reported in
the case of La$_{0.67}$Sr$_{0.33}$MnO$_{3}$/SrRuO$_3$
bilayers\cite{ke} where the M(H) loop shifts in the same direction
as that of the biasing field with an H$_{ex}$ $\approx$ 98 Oe.
Although the EB in our case is slightly smaller ($\approx$ 68 Oe),
we believe the LSMO(AF) is better suited than SRO for exchange
bias purposes because of the later's ferromagnetic character and a
large uniaxial magnetocrystalline anisotropy. For the trilayer
sample-T (Panels c $\&$ d of Fig. 3) the coercive field H$_C$ of
cobalt remains fixed at $\approx$ 42 Oe where as the H$_C$ of
LSMO(FM) can be shifted back and forth by exchange bias-controlled
pinning field. This difference in the coercivity results in a
distinct step in the M-H loop, thus making it a very interesting
system from application point of view. Moreover the coercivity of
LSMO(FM) can be further engineered by tuning the field used for
cooling and the relative thickness of the FM and AF layers. Here
it is important to point out that the coercive field of thin
epitaxial LSMO(FM) films deposited on (100)STO is comparable to
the H$_C$ of Cobalt\cite{senapati}. Hence exchange bias of
LSMO(FM)/Co bilayer with LSMO(AF) is necessary to realize
coercivity contrast. Also the biasing direction can be reset by
warming the sample back up to room temperature and cooling again
with the field reoriented.

The negative H$_{ex}$ with respect to reference cooling direction
can be understood by ferromagnetic coupling between uncompensated
moments in the (001) plane of LSMO(AF) and the moments of LSMO(FM)
so that the latter are frozen in the direction of the applied
field and therefore a bigger force or stronger external field in
opposite direction is required to overcome this coupling. It is
well known that for FM/AF bilayer the magnitude of H$_{ex}$ is
given by\cite{nogues}; $\left| {H_{ex} } \right|$ = J/M$_F$t$_F$,
where J is the FM-AF interfacial exchange coupling energy, M$_F$
and t$_{F}$ are the magnetization per unit volume and thickness of
the FM layer respectively. Substituting M$_F$ $\approx$ 400 emu/cc
extracted from the data of Fig. 3(a), $\left| {H_{ex} } \right| $
$\approx$ 68 Oe , and t$_{F}$ $\approx$ 50 nm, the exchange
coupling energy J comes out to be $\approx$ 0.13 erg/cm$^2$ at 10
K. The value of J for this bilayer system is comparable to J of
conventional metallic bilayers like NiMn/FeNi (J $\approx$ 0.27
erg/cm$^2$)\cite{lin,nogues}, Co/IrMn (J $\approx$ 0.12
erg/cm$^2$)\cite{julio} and MnPt/FeNi (J $\approx$ 0.03 erg/cm$^2$
)\cite{farrow} etc.

The EB effect also results in enhancement of coercivity of
LSMO(FM) layer which can be used for bringing coercivity contrast
between LSMO(FM) and cobalt in LSMO/STO/Co tunnel junctions. A
typical field dependence of magnetization and resistance for a
LSMO(AF)/LSMO(FM)/STO/Co tunnel junction measured at 30 K is shown
in the Fig. 4(a) and (b) respectively\cite{mudujn}. A step-like
inverse TMR results because of the antiparallel configuration
between LSMO(FM) and Co layers in the field range of 50-150 Oe.
The resistance step also found to match well with the
magnetization step as shown in Fig. 4(a). The TMR of these
junctions is $\approx$ 6 $\%$ at 10 K. De Teresa $\emph{et
al.}$\cite{teresa} have reported a strong bias dependent TMR in
LSMO/STO/Co with a peak value of $\approx$ 30 $\%$ at 10 K in
junctions of much smaller area. We are currently investigating the
bias and area dependence of magnetoresistance in our junctions.

In summary we have demonstrated that antiferromagnetic
La$_{0.45}$Sr$_{0.55}$MnO$_{3}$ can provide a robust exchange bias
to FM La$_{0.67}$Sr$_{0.33}$MnO$_{3}$. The advantage of this
material are its identical chemistry and close lattice match with
the ferromagnetic LSMO(FM). The direction of exchange bias is
shown to be controlled by the field used for cooling. The shift in
the hysteresis loop was also used to bring coercivity contrast in
La$_{0.45}$Sr$_{0.55}$MnO$_{3}$/La$_{0.67}$Sr$_{0.33}$MnO$_{3}$/SrTiO$_{3}$/Co
heterostructures. Such shift has been used to realize sharp
magnetic field controlled switching of conductance in tunnel
junctions made of La$_{0.67}$Sr$_{0.33}$MnO$_{3}$/SrTiO$_{3}$/Co.
% End main body
\begin{acknowledgements}
This research has been supported by grants from Board of Research
in Nuclear Sciences (BRNS) and Department of Information
Technology (DIT), Govt. of India.  P. K. Muduli acknowledges
financial support from the Council for Scientific and Industrial
Research (CSIR), Govt. of India.
\end{acknowledgements}

\clearpage
\begin{figure}[h]
\begin{center}
%\vskip -1.5cm
\abovecaptionskip -10cm
\includegraphics [width=12 cm]{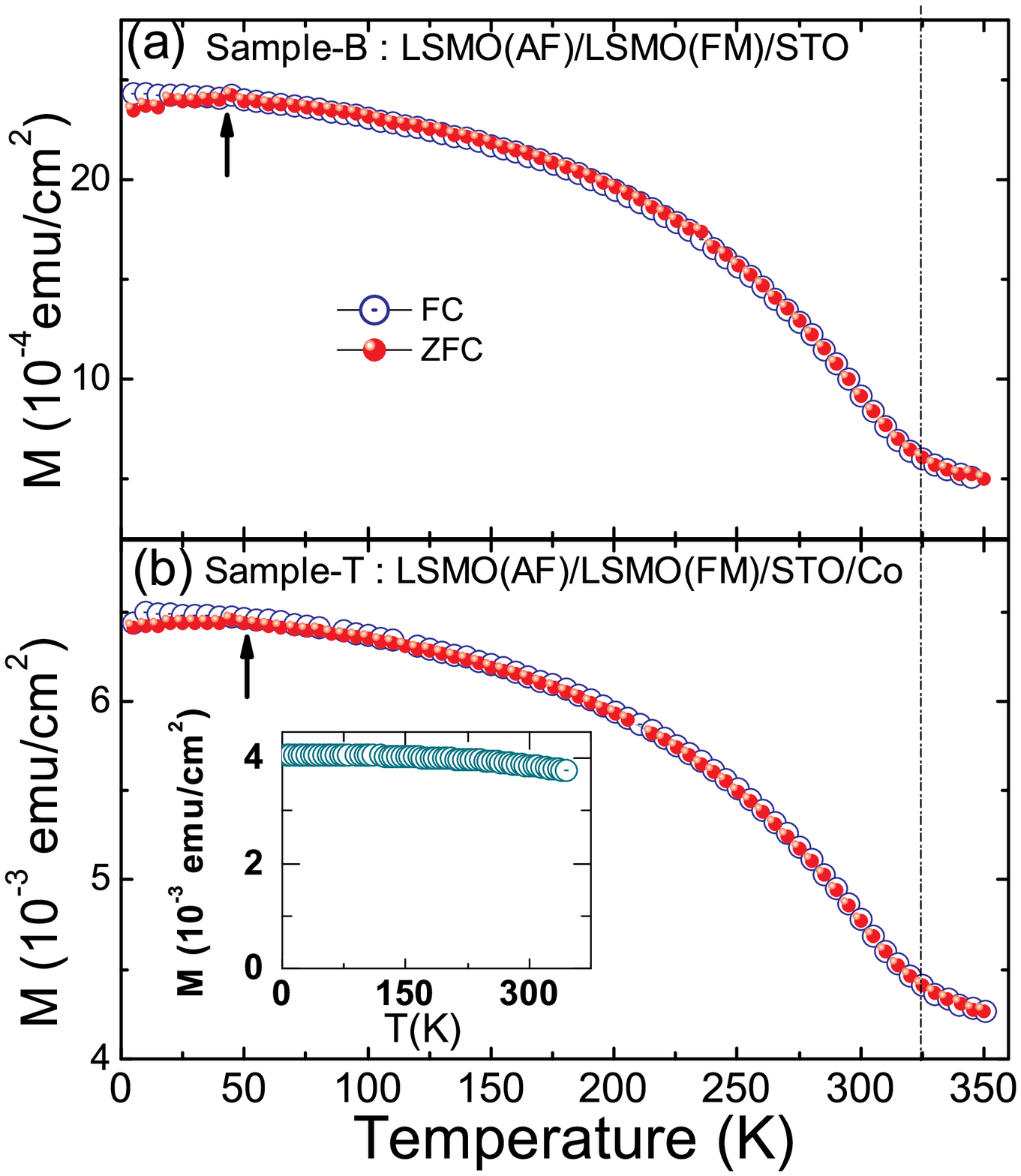}
\end{center}
\caption{\label{fig1} Temperature dependence of  field-cooled (FC)
(open circle) and zero-field-cooled (ZFC)(closed circle)
magnetization of Sample-B (a) and Sample-T (b)  measured in 1500
Oe applied in the plane of the film. The inset in (b) shows the
temperature dependence of FC magnetization of the cobalt layer
determined by subtracting the FC magnetization values of sample-B
and Sample-T.}
\end{figure}
\clearpage
\begin{figure}[h]
\begin{center}
%\vskip -1.5cm
\abovecaptionskip -10cm
\includegraphics [width=12cm]{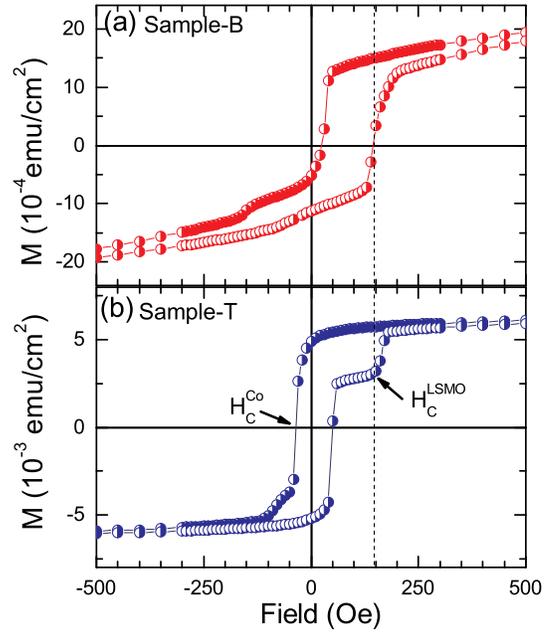}%
\end{center}
\caption{\label{fig1} Field dependence of in-plane magnetization
of Sample-B (a)and Sample-T (b) measured at 10 K after cooling the
sample in zero-field.}
\end{figure}
\clearpage
\begin{figure}[h]
\begin{center}
%\vskip -1.5cm
\abovecaptionskip -10cm
\includegraphics [width=12cm]{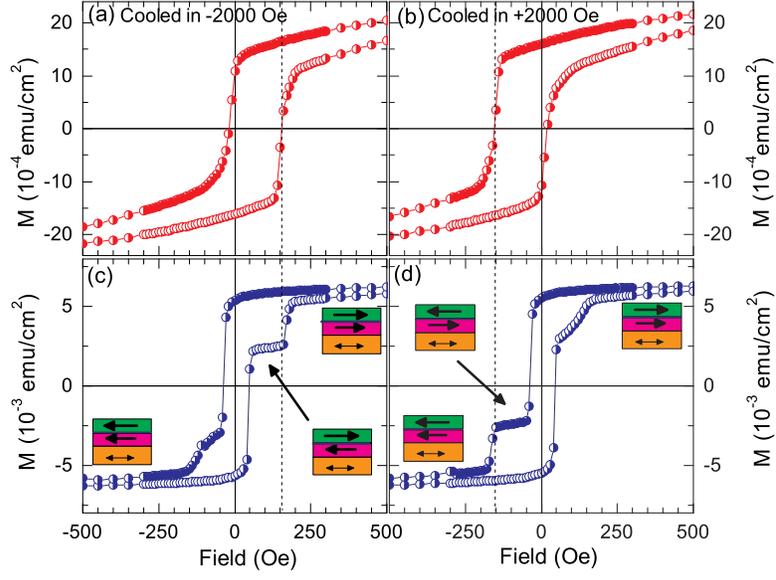}%
\end{center}
\caption{\label{fig1} Field dependence of magnetization of
Sample-B (a $\&$ b)  and Sample-T (c $\&$ d) measured at 10 K
after cooling the sample in $\pm$ 2000 Oe field. The field axis in
case of Sample-T can be divided into three zones according to
relative orientation of magnetization of each layers as shown in
boxes.}
\end{figure}

\clearpage
\begin{figure}[h]
\begin{center}
%\vskip -1.5cm
\abovecaptionskip -10cm
\includegraphics [width=12cm]{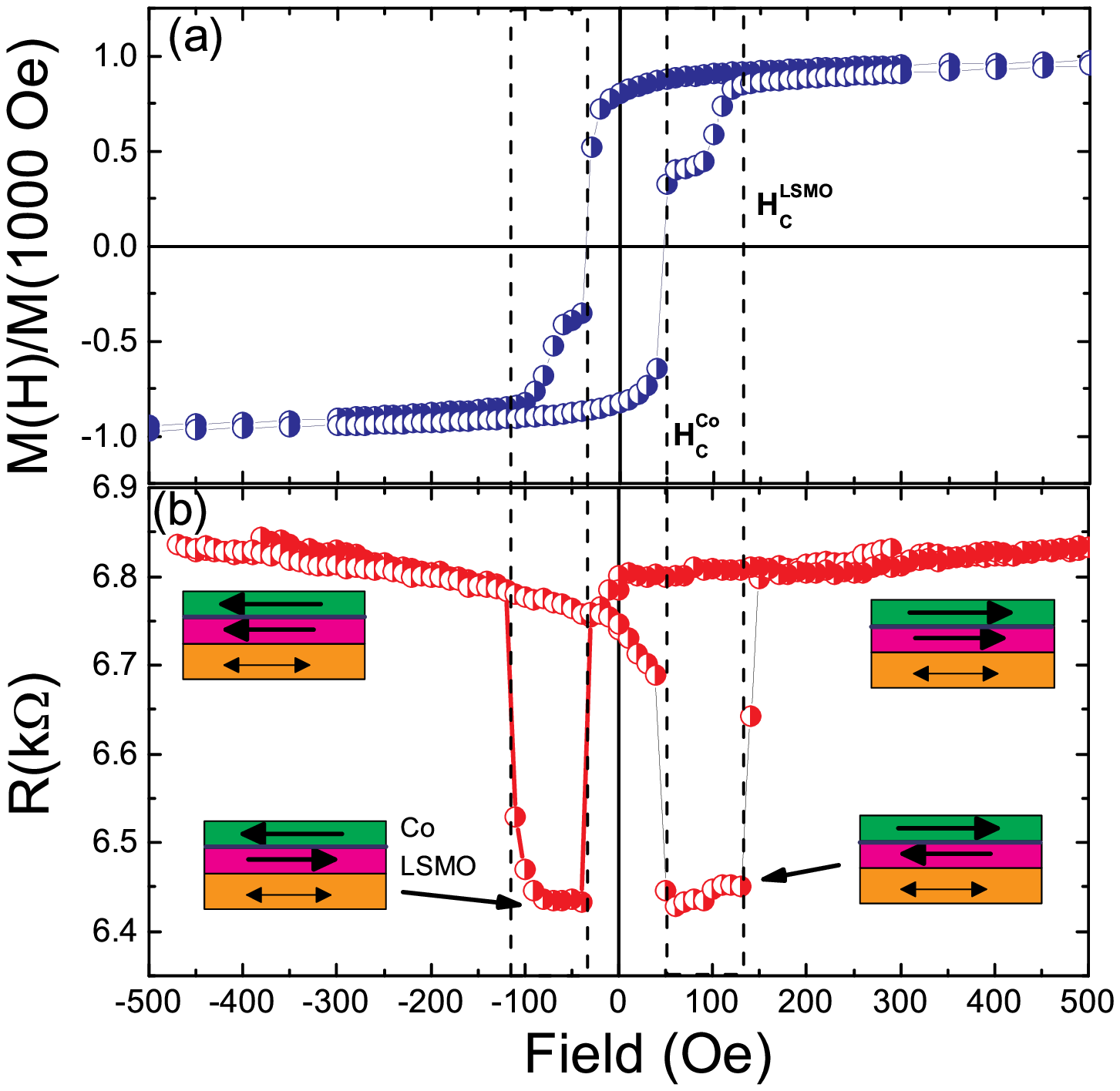}%
\end{center}
\caption{\label{fig1} Field dependence of (a) magnetization and
(b) resistance of a
LSMO(AF)(80nm)/LSMO(FM)(50nm)/STO(3nm)/Co(50nm) tunnel junction
measured at 30 K. Low resistance state between H$_C$$^{Co}$ and
H$_C$$^{LSMO}$ is due to inverse TMR observed in LSMO/STO/Co
tunnel junctions.}
\end{figure}


\begin{thebibliography}{100}
\bibitem{mudujn}
200 nm Y$_2$O$_3$ was deposited through a shadow mask by e-beam
evaporation on Sample-T leaving a 25 $\times$ 4000 $\mu$m$^2$
channel uncovered. Cross stripe of 50 nm thick and 25 $\mu$m wide
cobalt film were deposited by e-beam evaporation to create
junctions of area 25 $\times$ 25 $\mu$m$^2$. Base contact was made
by Ar$^+$ ion milling of SrTiO$_3$ layer at the end of sample.
These stripes were finally covered with 150 nm thick Ag-capping
layer.
\end{thebibliography}
\end{document}